\documentclass[preprint,5p,twocolumn]{elsarticle}

\usepackage{graphicx}
\usepackage{subcaption}
\usepackage{lipsum}
\usepackage{booktabs}
\usepackage{tabularx}
\usepackage{tikz}
\usepackage{pgfplots}
\usetikzlibrary{positioning,shapes,trees,arrows,graphs,decorations,calc}
\usepackage{url}

\usepackage{amssymb}
\usepackage{amsmath}
\usepackage{hyperref}
\usepackage{xcolor}

\usepackage{placeins} 
\usepackage[normalem]{ulem}
\usepackage{xspace}

\newcommand{\eq}[1]{eq.~\eqref{eq:#1}}
\newcommand{\eqs}[2]{eqs.~\eqref{eq:#1} and \eqref{eq:#2}}
\renewcommand{\sec}[1]{sec.~\ref{sec:#1}}

\newcommand{\fig}[1]{Fig.~\ref{fig:#1}}

\newcommand{\app}[1]{~\ref{app:#1}}

\newcommand{\ord}[1]{\mathcal{O}(#1)}

\newcommand{\df}{\mathrm{d}}

\newcommand{\Tau}{\mathcal{T}}

\newcommand{\GeV}{\,\mathrm{GeV}}

\newcommand{\nn}{\nonumber}

\newcommand{\cP}{\mathcal{P}}

\newcommand{\cut}{\mathrm{cut}}

\newcommand{\NLO}{\mathrm{NLO}}

\newcommand{\one}{{(1)}}

\newcommand{\dsigMC}{\df\sigma^\textsc{mc}}
\newcommand{\dsigtildeMCz}{\widetilde{\df\sigma^\textsc{mc}_0}}

\newcommand{\geneva}{\textsc{Geneva}\xspace}

\newcommand{\powheg}{\textsc{Powheg}\xspace}

\newcommand{\pythiaEight}{\textsc{Pythia8}\xspace}

\newcommand{\rivet}{\textsc{Rivet}\xspace}

\newcommand{\openloopsTwo}{\textsc{OpenLoops2}\xspace}
\newcommand{\minnlops}{\textsc{MiNNLO}$_{\rm PS}$\xspace}

\renewcommand{\matrix}{\textsc{Matrix}\xspace}
\newcommand{\munich}{\textsc{Munich}\xspace}

\def\thetaPSiso{\ensuremath{\Theta^{\mathrm{PS}}}\xspace}

\def\thetaProj{\ensuremath{\Theta^{\mathrm{proj}}\xspace}}

\def\thetaBarProj{\ensuremath{\overline{\Theta}^{\mathrm{proj}}\xspace}}

\biboptions{comma,compress}

\newcounter{bla}

\newcommand{\rescaletwoplots}{0.49\textwidth}
\newcommand{\hspacebetweentwoplots}{1em}

\newcommand{\rescalethreeplots}{0.32\textwidth}
\newcommand{\hspacebetweenthreeplots}{1em}

\newcommand{\spaceabovefigurecaption}{-1ex}

\makeatletter
\g@addto@macro\bfseries{\boldmath}
\makeatother

\begin{document}

\begin{frontmatter}

\title{$W\gamma$ production at NNLO+PS accuracy in \geneva}

\author[label1]{Thomas Cridge}
\author[label2]{Matthew A.~Lim\corref{author}}
\author[label2]{Riccardo Nagar}

\cortext[author] {Corresponding author.\\\textit{E-mail address: matthew.lim@unimib.it}}

\address[label1]{Department of Physics and Astronomy, University College London,
  London, WC1E 6BT, UK}
\address[label2]{Universit\`a degli Studi di Milano-Bicocca \& INFN
  Sezione di Milano-Bicocca, Piazza della Scienza 3, Milano 20126,
  Italy}

\date{\today}

\begin{abstract}
  We present an event generator for the process \mbox{$pp\to \ell\nu_\ell\gamma$}
  at next-to-next-to-leading order (NNLO) in QCD and matched to the \pythiaEight parton shower.
  The calculation makes use of the \geneva framework,
  which combines a resummed calculation obtained via Soft-Collinear Effective Theory (SCET)
  with the fixed-order result.
  We validate the NNLO accuracy of our results by comparing predictions for inclusive quantities
  with an independent fixed-order calculation,
  and then present the resummed $0$-jettiness spectrum at next-to-next-to-leading logarithmic (NNLL$'$) accuracy.
  Finally, we compare our predictions against data collected by the ATLAS experiment
  at the Large Hadron Collider during its 7 TeV run, finding good agreement.
\end{abstract}

\end{frontmatter}

\section{Introduction}
\label{sec:intro}
The process \mbox{$pp\to \ell\nu_\ell\gamma$} is particularly interesting among diboson processes measurable at the Large Hadron Collider (LHC). Due to the presence of a radiation zero at leading order\footnote{The tree-level $W\gamma$ diagram  featuring the non-Abelian coupling vanishes exactly at a centre-of-mass scattering angle of $\cos \theta^\ast = 1 / 3$, suppressing the leading order partonic cross section~\cite{Mikaelian:1979nr}.}, the next-to-leading order (NLO) corrections in the strong coupling are artificially large. This motivates the inclusion of higher order (next-to-next-to-leading, or NNLO) corrections in theoretical predictions for this process. It is also sensitive to the value of the non-Abelian electroweak gauge coupling, making it important for studies which seek to stress-test the gauge structure of the Standard Model (SM) and probe the effects of additional higher-dimensional operators in the SM effective field theory (SMEFT).

The $W\gamma$ final state was originally observed in $p\bar{p}$ collisions at the Tevatron~\cite{D0:1995mca,Acosta:2004it,Abazov:2005ni,Abazov:2011rk}, and has since been measured by both the ATLAS~\cite{Aad:2011tc,Aad:2012mr,Aad:2013izg} and CMS~\cite{Chatrchyan:2011rr,Chatrchyan:2013fya,Sirunyan:2021zud} collaborations at centre-of-mass energies of 7 and 13 TeV. These measurements have been used to constrain anomalous $WW\gamma$ couplings and, in the case of Ref.~\cite{Sirunyan:2021zud}, to set limits on SMEFT Wilson coefficients.

Fixed order calculations for this process at NNLO accuracy first appeared in Ref.~\cite{Grazzini:2015nwa}. These were made publicly available in the \matrix program~\cite{Grazzini:2017mhc}, which employs the $q_T$ slicing method to regularise infrared divergences. Electroweak corrections to this process at NLO are also known~\cite{Accomando:2001fn}, and have been combined with NLO~\cite{Denner:2014bna} and NNLO~\cite{Kallweit:2019zez} QCD corrections. Very recently, a new calculation of the process at NNLO QCD accuracy appeared~\cite{Campbell:2021mlr}, which utilised the $N$-jettiness slicing method and analytic expressions for the multi-loop matrix elements. An NLO+PS event generator was also implemented in the \powheg framework \cite{Barze:2014zba}.

There has been much recent activity in the field of matching calculations at NNLO to parton showers (PS), with four main approaches extant~\cite{Alioli:2015toa,Alioli:2021qbf,Hamilton:2013fea,Monni:2019whf,Hoeche:2014aia,Hoche:2014dla,Hu:2021rkt}. Event generators at NNLO+PS accuracy for several diboson processes are already available -- in the \geneva framework, diphoton production~\cite{Alioli:2020qrd}, $ZZ$ production~\cite{Alioli:2021egp}, and Higgsstrahlung~\cite{Alioli:2019qzz} have all been implemented, while processes such as $WW$ production~\cite{Re:2018vac,Lombardi:2021rvg} and $Z\gamma$~\cite{Lombardi:2020wju} are also available at NNLO+PS via the \minnlops method.

In this work, we present an event generator at NNLO+PS accuracy for the $W\gamma$ process. The \geneva approach, which we follow in this work, relies on a consistent matching of calculations resummed in a jet resolution variable at a high logarithmic accuracy with fixed order predictions at NNLO. The events thus produced are then fed to a parton shower program, which creates final states of high multiplicity and can be compared to experimental data. The framework is fully general, in the sense that it does not depend on a particular choice of resolution variable or a particular resummation formalism, nor is it restricted in its application to a specific class of processes. For this case, we make use of the $0$-jettiness resolution variable and its resummation in Soft-Collinear Effective Theory (SCET), which have been successfully employed in previous \geneva calculations. We shower the partonic events with \pythiaEight, which also simulates hadronisation.

The Letter is organised as follows. In \sec{proc} we define the process and provide a brief recap of the \geneva formulae, paying particular attention to the caveats that are needed for processes such as ours which are divergent at Born level. In \sec{results} we present a validation against the \matrix result at NNLO and compare our predictions against ATLAS data. We conclude in \sec{conc}. In \app{power_corrections}, we discuss the impact of missing power corrections on our results and the reweighting procedure that we use to limit this.

\section{Process definition}
\label{sec:proc}

We apply the \geneva formalism to the process \mbox{$pp\to \ell\nu_\ell\gamma + X$}.
The presence of a photon in the final state requires the introduction
of criteria which ensure that the photon is produced in the hard interaction, rather than being the result of fragmentation from a quark or gluon. To this end, we
employ the smooth-cone isolation procedure introduced by Frixione~\cite{Frixione:1998jh} which limits the amount of transverse hadronic energy around the photon in
an infrared-safe manner. Explicitly, the procedure requires that a continuous series of sub-cones with radius $r\leq R_{\mathrm{iso}}$ satisfy the criterion $ E^{\textrm{had}}_T(r) \leq E^{\textrm{max}}_T \, \chi(r;R_{\textrm{iso}})$,
where a standard choice for the isolation function is
\begin{align}
\chi(r;R_{\text{iso}}) = \left(\frac{1-\cos r}{1-\cos{R_{\text{iso}}}}\right)^{n}\!,
\end{align}
though other choices are possible. In this way, hadronic activity is smoothly reduced as one approaches the photon direction until at $r=0$ it is completely absent.

The smooth-cone isolation procedure has several theoretical advantages such as infrared safety. It is, however, somewhat at odds with the approach taken in collider experiments, where the discretised
nature of the detector means that the smooth sub-cone concept is not easily realised in practice, and it is instead common to consider a cone of fixed size $R_{\mathrm{iso}}$. For this reason,
Refs.~\cite{Chen:2019zmr,Siegert:2016bre} investigated the possibility of using a hybrid isolation procedure in theoretical predictions. In this approach, events are generated subject to a smooth-cone isolation with
only loose cuts being placed (i.e. a small $R_{\mathrm{iso}}$ is used) and subsequently tighter cuts are imposed at analysis level using a cone of fixed size. In this way one combines
the advantages of both approaches and makes the direct comparison to experimental data possible. We make use of this approach in our comparison to experimental data in \sec{results}.

In addition to the photon isolation, it is also necessary to impose other process-defining cuts to ensure that the Born-level cross section is free of QED divergences. Specifically,
one needs to separate in some way the photon from the initial-state partons and also from the charged lepton produced in the $W$ decay. We achieve these separations by placing
requirements on the transverse momentum of the photon with respect to the beam direction $p_T^\gamma\geq p_{T,\cut}^\gamma$ and on the $\Delta R\equiv \sqrt{(\Delta \phi)^2 + (\Delta y)^2}$ of the lepton-photon pair $\Delta R^{\ell\gamma} \geq \Delta R_{\cut}^{\ell\gamma}$.

The Monte Carlo (MC) cross sections of various multiplicity in \geneva are defined to correspond directly to the weights of physical and IR-finite events at a given perturbative accuracy. This
is achieved by partitioning the phase space into regions with different numbers of resolved emissions classified according to some $N$-jet resolution variables $\Tau_N$, and then mapping IR-divergent final
states with $M \ge N$ partons to IR-finite final states with $N$ partonic jets. This ensures the cancellation of IR singularities on an event-by-event basis. The implication is that
the \geneva MC cross sections $\dsigMC_N$ receive contributions from configurations with both $N$ and $M$ partons, where the additional emissions lie below a cut
$\Tau_N^\cut$ and are considered unresolved. By resumming the resolution parameters at a high logarithmic accuracy, the dependence on the partition boundaries $\Tau_N^\cut$ is then substantially mitigated.

Although the \geneva method does not rely on any specific choice for the resolution variables $\Tau_N$ (and indeed several different choices are possible), thus far most implementations have relied
on the $0-$ and $1-$jettiness variables, defined as~\cite{Stewart:2010tn}
\begin{align} \label{eq:TauNdef}
\Tau_N = \sum_k \min \Bigl\{ \hat q_a \cdot p_k, \hat q_b \cdot p_k, \hat q_1 \cdot p_k, \ldots , \hat q_N \cdot p_k \Bigr\}
\,,\end{align}
with $N=0$ or $1$,  $q_{a,b}$ representing the beam directions, $q_{1 \dots N}$ any final-state massless directions that minimise $\Tau_N$, and $p_k$ the final-state partonic momenta. These separate the $0$- and $1$-jet exclusive from the $2$-jet inclusive cross sections.

The \geneva formulae, which have been presented in full in several papers~\cite{Alioli:2015toa,Alioli:2019qzz}, are slightly complicated by the presence of process-defining cuts at Born level. Since
the additional complications which arise have been discussed in detail in Ref.~\cite{Alioli:2020qrd}, in this Letter we simply state the final results. The differential cross sections in the presence of process-defining cuts $\thetaPSiso(\Phi_N)$ are given by
\begin{align} \label{eq:0full}
&\frac{\dsigMC_0}{\df\Phi_0}(\Tau_0^\cut) = \\
&\qquad\quad\frac{\df\sigma^{\rm NNLL'}}{\df\Phi_0}(\Tau_0^\cut) - \biggl[\frac{\df\sigma^{\rm NNLL'}}{\df\Phi_{0}}(\Tau_0^\cut) \biggr]_{\rm NNLO_0} \nn \\
&\qquad +(B_0+V_0+W_0)(\Phi_0) \, \thetaPSiso(\Phi_0) \nn \\
&\qquad +  \int \frac{\mathrm{d} \Phi_1}{\mathrm{d} \Phi_0} (B_1 + V_1)(\Phi_1)\, \thetaPSiso (\Phi_1) \nn \\
&\qquad\qquad\times\,\thetaProj(\widetilde{\Phi}_0)\,\theta\big( \Tau_0(\Phi_1)< \Tau_0^{\mathrm{cut}}\big) \nn \\
&\qquad+  \int \frac{\mathrm{d} \Phi_2}{\mathrm{d} \Phi_0} \,B_2 (\Phi_2)\, \thetaPSiso (\Phi_2) \, \theta\big( \Tau_0(\Phi_2)< \Tau_0^{\mathrm{cut}}\big)\,, \nn
\end{align}
\begin{align} \label{eq:1belowtau0}
&\frac{\dsigMC_{1}}{\df\Phi_{1}} (\Tau_0 \le \Tau_0^\cut; \Tau_{1}^\cut) = \\
&\quad B_1\,(\Phi_1)\, \thetaPSiso (\Phi_1) \,\,  \big[ \overline{\Theta}_{\mathrm{iso}}^{\mathrm{proj}}(\widetilde{\Phi}_0)  + \overline{\Theta}^{\mathrm{FKS}}_{\mathrm{map}}(\Phi_1)  \big] \theta(\Tau_0<\Tau^{\mathrm{cut}}_0)\,, \nn
\end{align}
\begin{align} \label{eq:1masterfull}
&\frac{\dsigMC_{1}}{\df\Phi_{1}} (\Tau_0 > \Tau_0^\cut; \Tau_{1}^\cut)= \\
&\thetaPSiso (\Phi_1) \,\Bigg\{\Bigg[ \frac{\df\sigma^{\rm NNLL'}}{\df\Phi_0\df\Tau_0}-\frac{\df\sigma^{\rm NNLL'}}{\df\Phi_0\df\Tau_0}\bigg\vert_{\NLO_1}\,\Bigg]\, \cP(\Phi_1)\, \thetaProj(\widetilde{\Phi}_0) \nn \\
&\quad+ \big[B_1 + V_1^C\big](\Phi_1)  \Bigg\} \, U_1(\Phi_1, \Tau_1^\cut)\, \theta(\Tau_0 > \Tau_0^\cut) \nn \\
&+\int\ \bigg[\frac{\df\Phi_{2}}{\df\Phi^\Tau_1}\,B_{2}(\Phi_2)\, \thetaPSiso (\Phi_2)\,\thetaProj (\widetilde{\Phi}_1)\, \theta\!\left(\Tau_0(\Phi_2) > \Tau_0^\cut\right) \nn \\
&\quad\times\,\theta(\Tau_{1} < \Tau_1^\cut) - \frac{\df\Phi_2}{\df \Phi^C_1}\, C_{2}(\Phi_{2})\,\thetaPSiso (\Phi_1) \, \theta(\Tau_0 > \Tau_0^\cut) \bigg] \nn \\
&- B_1(\Phi_1)\, U_1^\one(\Phi_1, \Tau_1^\cut)\,\thetaPSiso (\Phi_1)\, \theta(\Tau_0 > \Tau_0^\cut)\,, \nn
\end{align}
\begin{align} \label{eq:2belowtau1}
&\frac{\dsigMC_{\geq 2}}{\df\Phi_{2}} (\Tau_0 > \Tau_0^\cut, \Tau_{1} \le \Tau_{1}^\cut) = \\
&\qquad B_2(\Phi_2)\, \thetaPSiso (\Phi_2)\, \Big[\thetaBarProj(\widetilde{\Phi}_1)+\overline{\Theta}_{\mathrm{map}}^\Tau(\Phi_2)\Big] \nn \\
&\qquad\qquad\times \theta(\Tau_1 < \Tau_1^\cut)\, \theta\left(\Tau_0(\Phi_2) > \Tau_0^\cut\right)\,, \nn
\end{align}
\begin{align} \label{eq:2masterfull}
&\frac{\dsigMC_{\geq 2}}{\df\Phi_{2}} (\Tau_0 > \Tau_0^\cut, \Tau_{1}>\Tau_{1}^\cut) = \\
& \thetaPSiso (\Phi_2)\, \Bigg\{ \bigg[ \frac{\df\sigma^{\rm NNLL'}}{\df\Phi_0\df\Tau_0} - \frac{\df\sigma^{\rm NNLL'}}{\df\Phi_0\df\Tau_0}\bigg|_{\NLO_1}\bigg]\, \cP(\widetilde{\Phi}_1)\, \thetaProj(\widetilde{\Phi}_0)\, \nn \\
&\quad + (B_1 + V_1^C)(\widetilde{\Phi}_1)\Bigg\} \,  U_1'(\widetilde{\Phi}_1, \Tau_1)\, \theta(\Tau_0 > \Tau_0^\cut) \Big\vert_{\widetilde{\Phi}_1 = \Phi_1^\Tau\!(\Phi_2)} \nn\\
&\quad\quad\times\,\thetaProj(\widetilde{\Phi}_1) \cP(\Phi_2) \, \theta(\Tau_1 > \Tau_1^\cut)
\nn \\
&+ \thetaPSiso (\Phi_2)\, \Big\{ B_2(\Phi_2)\, \theta(\Tau_{1}>\Tau^{\mathrm{cut}}_{1})- B_1(\Phi_1^\Tau)\,U_1^{\one\prime}\!\big(\widetilde{\Phi}_1, \Tau_1\big)\nn\\
&\quad\times\,\cP(\Phi_2)\,\thetaProj(\widetilde{\Phi}_1)\, \Theta(\Tau_1 > \Tau_1^\cut)
\Big\}\, \theta\left(\Tau_0(\Phi_2) > \Tau_0^\cut\right)\,. \nn
\end{align}

The $B_j$, $V_j$ and $W_j$ are the $0$-, $1$- and $2$-loop
matrix elements for $j$ QCD partons in the final state; we denote by $\mathrm{N}^k\mathrm{LO}_n$ a quantity with $n$ additional partons in the final state
computed at $\mathrm{N}^k\mathrm{LO}$ accuracy.

The formulae above require one to evaluate the resummed and resummed-expanded terms on phase space points resulting
from a projection from a higher to a lower multiplicity. We denote such projected phase space points by $\widetilde{\Phi}_{N}$.
It is vital that these projected configurations also satisfy the process-defining
cuts. We denote by $\thetaProj(\widetilde{\Phi}_N)$ the set of restrictions acting on the higher dimensional
$\Phi_{N+1}$ phase space due to the cuts on $\widetilde{\Phi}_N$. In practice, this means that when
a term in the cross section, evaluated at a $\Phi_{N+1}$ phase space point, is multiplied by $\thetaProj(\widetilde{\Phi}_N)$,
we perform the projection $\Phi_{N+1}\to \widetilde{\Phi}_{N}$ and apply the cuts to $\widetilde{\Phi}_{N}$. If the projected point
fails the cuts, the initial $\Phi_{N+1}$ configuration is excluded.

We use the shorthand
\begin{align}
\label{eq:dPhiRatio}
 \frac{\df \Phi_{M}}{\df \Phi_N^{\cal O}}  = \df \Phi_{M} \, \delta[ \Phi_N - \Phi^{\cal O}_N(\Phi_M) ] \,\Theta^{\cal O}(\Phi_M)
\end{align}
to indicate an integration over the portion of the $\Phi_M$ phase space which can be reached from a fixed value of $\Phi_N$ while keeping some observable $\cal O$ also fixed, with $N \leq M$.
The $\Theta^{\cal O}(\Phi_M)$ term additionally limits the integration to
the phase space points belonging to the singular contribution for the
given observable $\cal O$.  For example, when generating $1$-body
events we use
\begin{equation} \label{eq:Phi1TauProj}
\frac{\df\Phi_2}{\df\Phi_1^\Tau} \equiv \df\Phi_2\,\delta[\widetilde{\Phi}_1 - \Phi^\Tau_1(\Phi_2)]\,\Theta^\Tau(\Phi_2)
\,,\end{equation}
where the map used by the $1 \to 2$ splitting has been constructed
to preserve $\Tau_0$, i.e.
\begin{equation} \label{eq:Tau0map}
\Tau_0(\Phi_1^\Tau(\Phi_2)) = \Tau_0(\Phi_2)
\,,\end{equation}
and $\Theta^\Tau(\Phi_2)$ guarantees that the $\Phi_2$ point is reached from a genuine QCD splitting of the $\Phi_1$ point.
The use of a $\Tau_0$-preserving mapping is necessary to ensure that the point-wise singular $\Tau_0$ dependence is alike among all terms in
\eqs{1masterfull}{2masterfull} and that the cancellation of said singular
terms is guaranteed on an event-by-event basis.\footnote{We remark that this differentiates our
  approach from the slicing method used in Ref.~\cite{Campbell:2021mlr} -- the use of a $\Tau_0$-preserving mapping ensures that our subtraction, though not fully local, results in point-wise cancellation of divergences in this single variable.}

The non-projectable regions of $\Phi_1$ and $\Phi_2$, on the other hand, belong to the cross sections in \eqs{1belowtau0}{2belowtau1}.
These events are entirely nonsingular in nature. We denote the constraints due to the choice of map by $\Theta_{\mathrm{map}}$~\footnote{We use the FKS
map~\cite{Frixione:2007vw} for the $\Phi_1 \to\widetilde{\Phi}_0$ projection and, as mentioned above, our own $\Tau_0$-preserving map for the
$\Phi_2 \to\widetilde{\Phi}_1$ projection.} and use $\thetaProj$ to indicate the restrictions due to both the isolation cuts and the flavour structure
of the underlying Born configuration (see e.g. Ref.~\cite{Alioli:2019qzz} for more details).

The term $V_1^C$ denotes the soft-virtual contribution of a standard NLO local subtraction:
\begin{align} \label{eq:FOFKS}
  V_1^C(\Phi_1) = V_1(\Phi_1)+\int\frac{\df\Phi_2}{\df \Phi_1^C}C_2(\Phi_2)\,,
\end{align}
 with $C_2$ a singular approximation of $B_2$; in practice we use the
 subtraction counterterms which we integrate over the radiation variables
 $\df\Phi_2 / \df \Phi_1^C$  using the singular limit $C$ of the phase space
 mapping.

 $U_1$ is an NLL Sudakov factor which resums large logarithms of $\Tau_1$, and $U_1'$ its derivative with respect to $\Tau_1$; the $\ord{\alpha_s}$
 expansions of these quantities are denoted by $U_1^\one$ and $U_1^{\one\prime}$ respectively.

We extend the differential dependence of the resummed terms from the $N$-jet to the $(N\!+\!1)$-jet phase space using a normalised splitting probability
$\mathcal{P}(\Phi_{N+1})$ which satisfies
\begin{align}
\label{eq:cPnorm}
\int \! \frac{\df\Phi_{N+1}}{\df \Phi_{N} \df \Tau_N} \, \cP(\Phi_{N+1}) = 1
\,.\end{align}
The two extra variables are chosen to be an energy ratio $z$ and an
azimuthal angle $\phi$. The functional forms of the $\cP(\Phi_{N+1})$ are
based on the Altarelli-Parisi splitting kernels, weighted by parton
distribution functions (PDFs) where appropriate.

\begin{figure*}[ht]
  \begin{subfigure}[b]{\rescalethreeplots}
    \includegraphics[width=\textwidth]{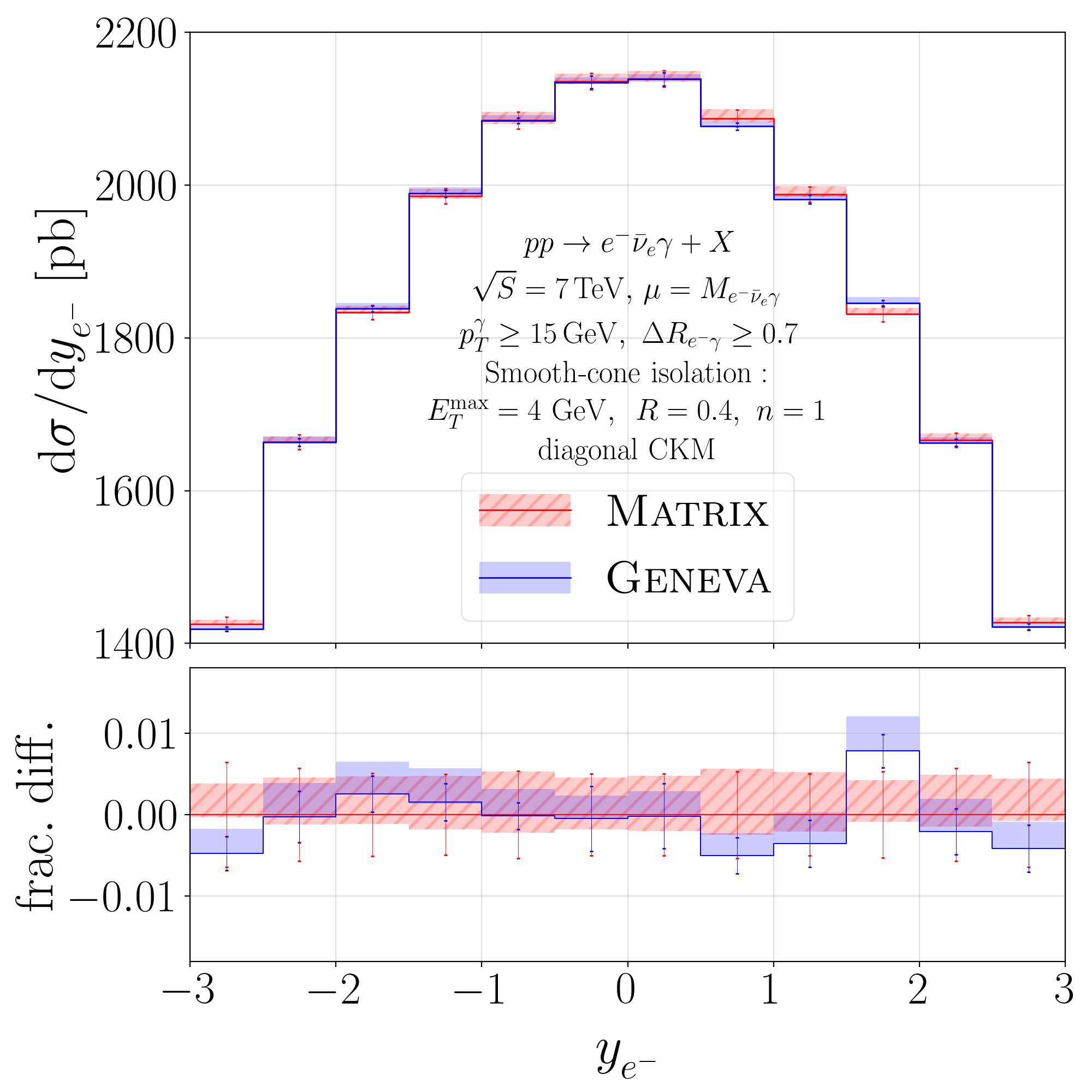}%
  \end{subfigure}
  \hspace*{\hspacebetweenthreeplots}
  \begin{subfigure}[b]{\rescalethreeplots}
    \includegraphics[width=\textwidth]{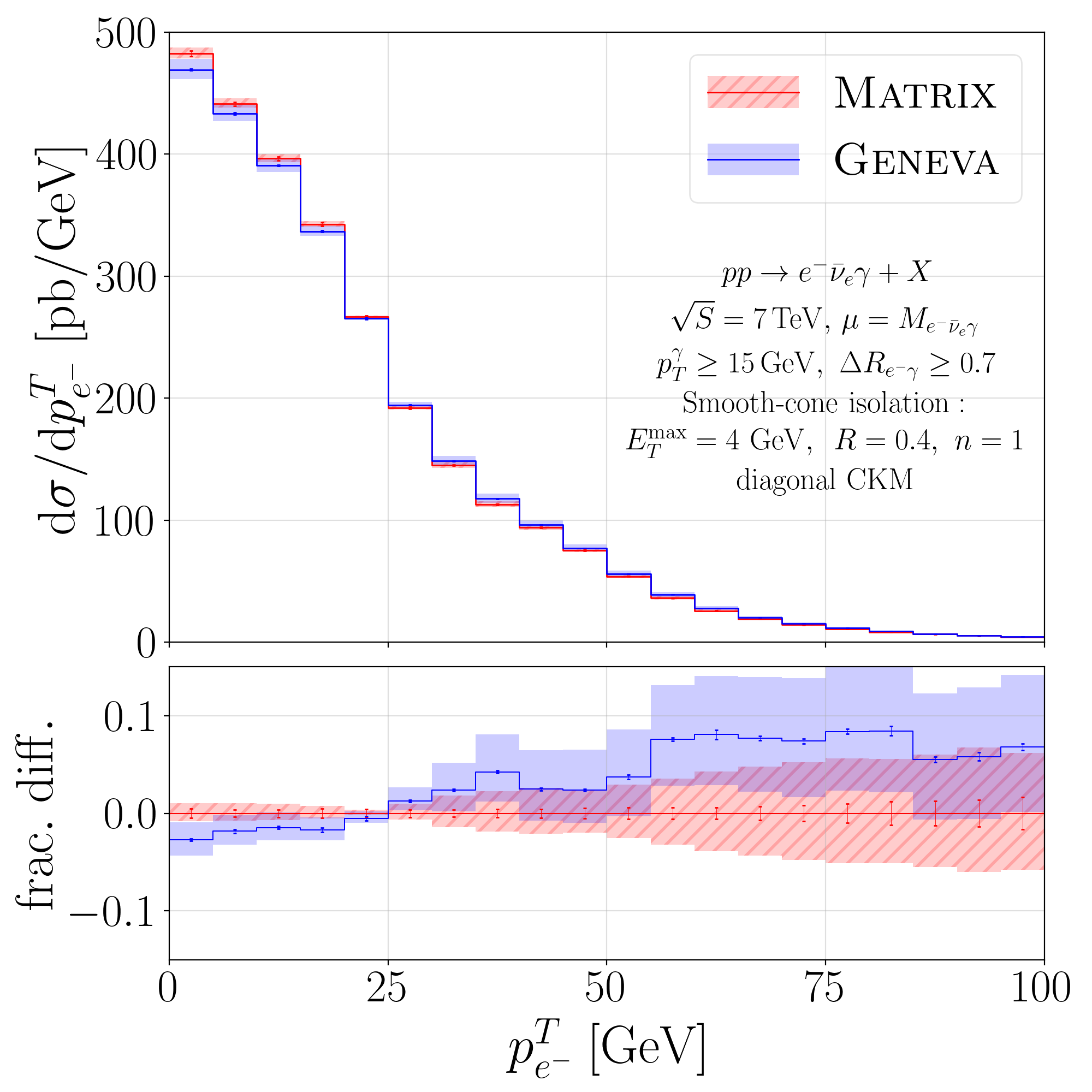}%
  \end{subfigure}
  \hspace*{\hspacebetweenthreeplots}
  \begin{subfigure}[b]{\rescalethreeplots}
    \includegraphics[width=\textwidth]{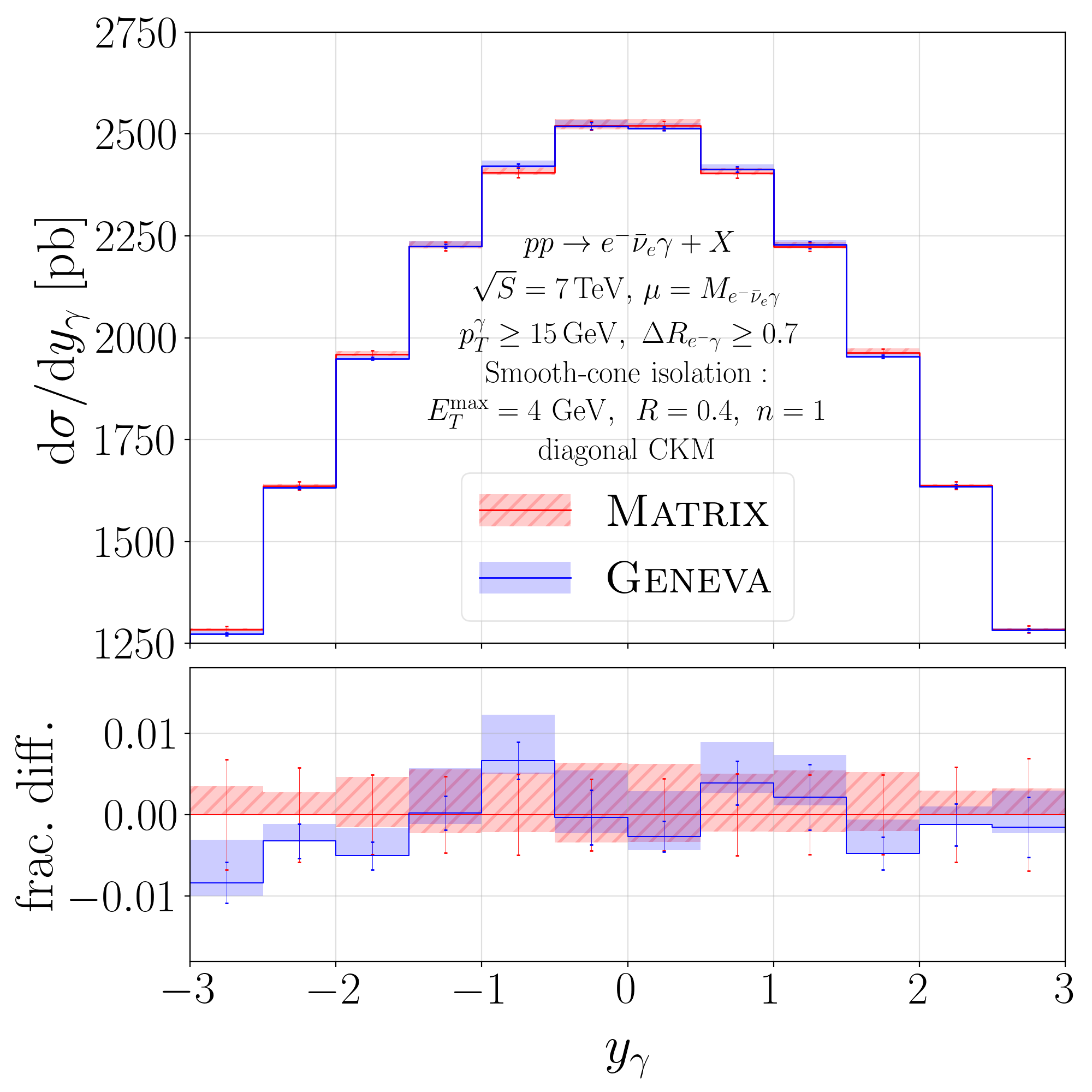}%
  \end{subfigure}
  \vspace{\spaceabovefigurecaption}
  \caption{Comparison between the \matrix and \geneva predictions at NNLO accuracy. We show the
    rapidity of the electron (left), the transverse momentum of the electron (centre) and the
    rapidity of the photon (right). Scale uncertainty bands include 3-point renormalisation and factorisation scale variations. Statistical errors connected to the Monte Carlo integration are shown as vertical error bars.
    \label{fig:gvavsmatrix}
  }
\end{figure*}

In principle, the \geneva method is compatible with any formalism which can provide the
resummed spectrum at NNLL$'$ accuracy. In practice, we obtain this from a factorisation theorem
derived in SCET and in the case of colour singlet production
write
\begin{align}\label{eq:factorization}
\frac{\textrm{d} \sigma^{\rm SCET}}{\textrm{d} \Phi_0 \textrm{d}
  \Tau_0}  = & \sum_{ij} H_{ij}(\Phi_0,\mu)\int\! \df t_a\, \df t_b \,B_i(t_a,x_a,\mu) \nn  \\ & \qquad
B_j(t_b,x_b,\mu) \, S(\Tau_0-\tfrac{t_a+t_b}{Q},\mu)\, ,
\end{align}
where the sum runs over all possible $q\bar{q}$ flavours. The cross section has been factorised into
hard $H_{ij}$, soft $S$ and beam $B_{i,j}$ functions which achieves a separation of scales: a judicious scale choice for each
component will therefore result in the absence of large logarithms. Resummation is then achieved via renormalisation group evolution
to a common scale $\mu$, viz.
\begin{align}
\label{eq:standardresum}
\frac{\textrm{d} \sigma^{\text{NNLL}^\prime}}{\textrm{d} \Phi_0 \textrm{d} \Tau_0} =\;& \sum_{ij} H_{ij}(\Phi_0,\mu_H) \, U_H(\mu_H,\mu)\, \nn \\ & \big\{\big[ B_i(t_a,x_a,\mu_B)\otimes U_B(\mu_B,\mu)\big]\, \nonumber \\
& \times \big[B_j(t_b,x_b,\mu_B)\otimes U_B(\mu_B,\mu)\big] \big\}\, \nn \\ &  \otimes \big[ S(\mu_s)\otimes U_S(\mu_S,\mu)\big]\, ,
\end{align}
where the $U_i$ denote the evolution kernels and the convolutions are written in a
schematic form. The cusp (non-cusp) anomalous dimensions up to $3$-($2$-)loop order needed for
the evolution at NNLL$'$ have been available for some time~\cite{Idilbi:2006dg,Becher:2006mr,Hornig:2011iu,Kang:2015moa,Gaunt:2015pea}. Other inputs to the factorisation theorem also appear in the literature -- the beam and soft functions have been computed in Refs.~\cite{Kelley:2011ng,Monni:2011gb,Gaunt:2014xga}, while the helicity amplitudes necessary for the extraction of the hard function were first computed in Ref.~\cite{Gehrmann:2011ab}.

The events thus produced are then fed to the \pythiaEight parton shower~\cite{Sjostrand:2007gs} which adds radiation to produce final states of a high multiplicity. This restores the radiation that was integrated over in the construction of the $0$- and $1$-jet cross sections, and adds additional particles to the inclusive $2$-jet bin. The consequence is that, while quantities which are inclusive over the radiation are expected to remain unmodified by the action of the shower, the distributions of more exclusive observables will be modified. We thus retain NNLO accuracy for inclusive quantities.

For further details on the topics discussed above, we refer the reader to Refs.~\cite{Alioli:2015toa,Alioli:2020qrd}.

\section{Results and comparison to LHC data}
\label{sec:results}
We begin by summarising the process-specific features of our implementation. At Born level, the process receives contributions from diagrams with three different resonance structures. Two of these are akin to the Drell-Yan process, but with the photon emitted either from the vector boson line or from the charged lepton, while the third is a $t$-channel process in which the photon is emitted from the quark line. In order to efficiently sample the phase spaces associated with these different resonance structures, we make use of the tunnel between \geneva and \munich \cite{munich}, which was first constructed in Ref.~\cite{Alioli:2021egp}. We rely on \openloopsTwo~\cite{Buccioni:2017yxi,Buccioni:2019sur} for the calculation of tree level and $1$-loop matrix elements. We make use of the $2$-loop hard function implemented for this process in \matrix~\cite{Grazzini:2015nwa} and use a diagonal CKM matrix, though this could in principle be trivially extended.\footnote{When performing our validation, we noticed discrepancies between the hard function at $1$-loop order as implemented in the public version 1.0.5 of \matrix and the results of \openloopsTwo, due to the use of an outdated version of the \texttt{hplog} package~\cite{Gehrmann:2001pz} in the former. We have corrected this in our implementation.}

In \fig{gvavsmatrix} we validate the NNLO accuracy of our predictions by comparison with \matrix. We consider the electronic decay channel of the $W^-$ boson at a $pp$ centre-of-mass energy of $\sqrt{S}=7\,\mathrm{TeV}$. We require a separation  $\Delta R^{e \gamma}>0.7$ between the electron and the photon, and place a transverse momentum cut on the photon $p_{T,\cut}^\gamma=15\,\mathrm{GeV}$. We employ a smooth-cone isolation procedure with parameters $E_T^{\mathrm{max}}=4\,\mathrm{GeV}$, $R=0.4$, $n=1$ and use the \texttt{MSHT20\_as118} sets~\cite{Bailey:2020ooq} from LHAPDF6
\cite{Buckley:2014ana}. We take $M_W=80.385\,\mathrm{GeV}$ and $\Gamma_W=2.085\,\mathrm{GeV}$ -- for the purposes of this comparison, we do not perform any resummation of the logarithms of $\Tau_1$.

In general, we observe a good agreement between the central values from \matrix and \geneva for the rapidity distributions. In the transverse momentum distribution, we notice distortions from the fixed order prediction similar to those observed in Ref.~\cite{Alioli:2020qrd} due to the omission of the kinematic dependence of nonsingular power corrections and the inclusion of higher order terms. Nevertheless, the two predictions agree within the scale variation bands. Examining the bands, we observe a difference in size between \matrix and \geneva which is particularly noticeable in the rapidity distribution of the electron. We remark that the reweighting procedure described in \app{power_corrections} is constructed to recover the value of the fixed order total cross section and its variations correctly, while the kinematic dependence of power corrections is still omitted. The effect on the scale bands is particularly noticeable in this case due to the accidentally small size of the scale uncertainties, which are a result of adopting a correlated variation with $\mu_R=\mu_F$ (see Ref.~\cite{Grazzini:2015nwa}).

\begin{figure}[t!]
    \includegraphics[width=0.45\textwidth]{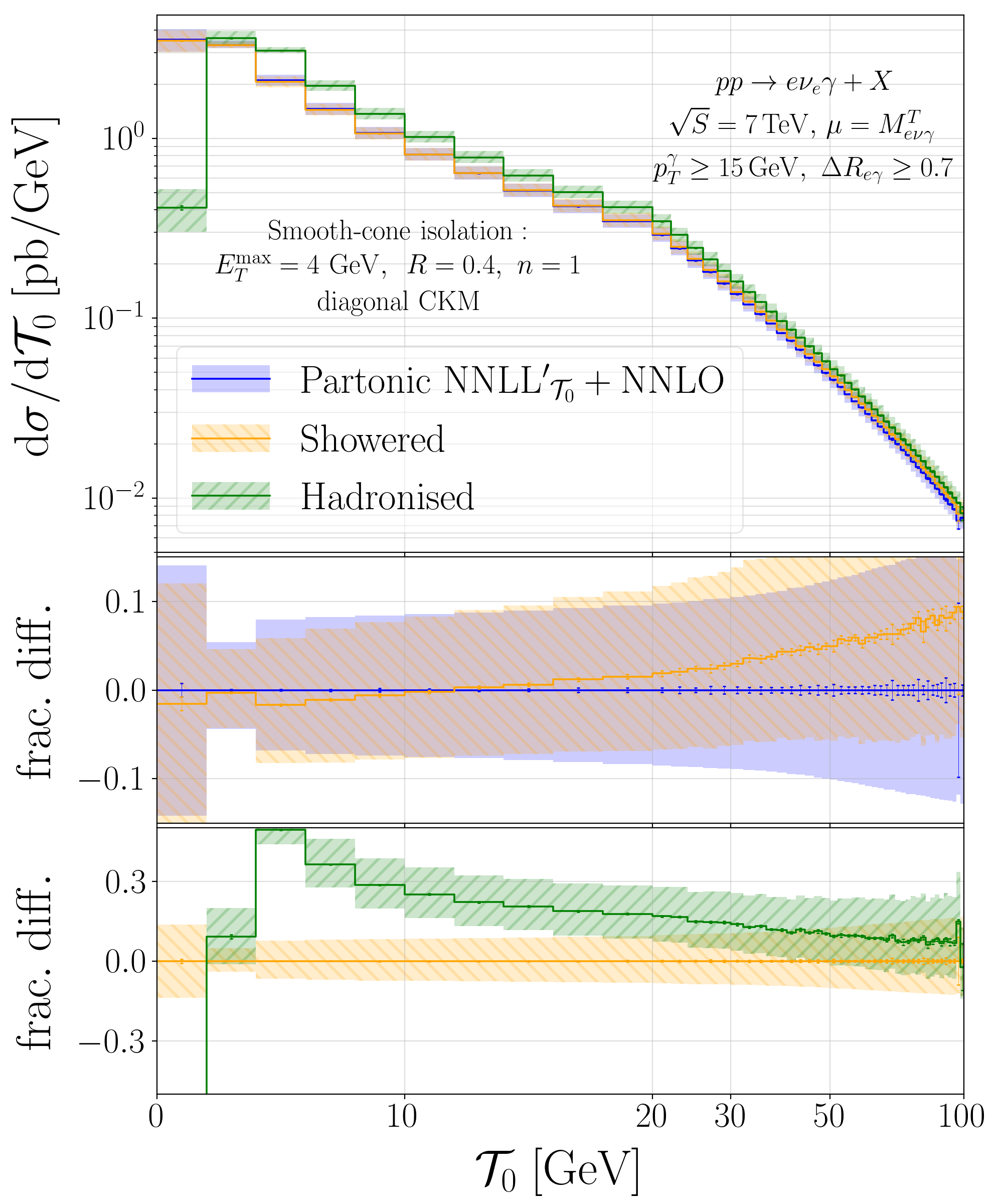}%
  \vspace{\spaceabovefigurecaption}
  \caption{The partonic $\Tau_0$ spectrum at NNLL$'$+NNLO$_0$ accuracy (blue) compared to results after showering (orange) and hadronisation (green).   The sum over both charges of the intermediate $W$ boson is shown.  A semi-logarithmic scale is used for $\Tau_0$, which is linear up to 20 GeV and logarithmic beyond.
    \label{fig:tau0spectrum}
  }
\end{figure}

The NNLL$'$+NNLO$_0$ accurate $\Tau_0$ spectrum at partonic level is shown in \fig{tau0spectrum}, where we also show the effects of showering and hadronisation with \pythiaEight. In order to simplify the analysis, we have deactivated both the QED part of the shower and also the simulation of multiparton interactions.  Comparing the partonic with the showered result, we note that the accuracy of the distribution is numerically well-preserved by the shower in the peak region (up to $\sim$ 10 GeV) where resummation effects are most important. The effect of hadronisation is to shift the peak of the distribution to higher values of $\Tau_0$ -- its impact is lessened in the tail of the distribution, where nonperturbative effects become increasingly less relevant.

\begin{figure*}[ht]
  \begin{subfigure}[b]{\rescaletwoplots}
    \includegraphics[width=\textwidth]{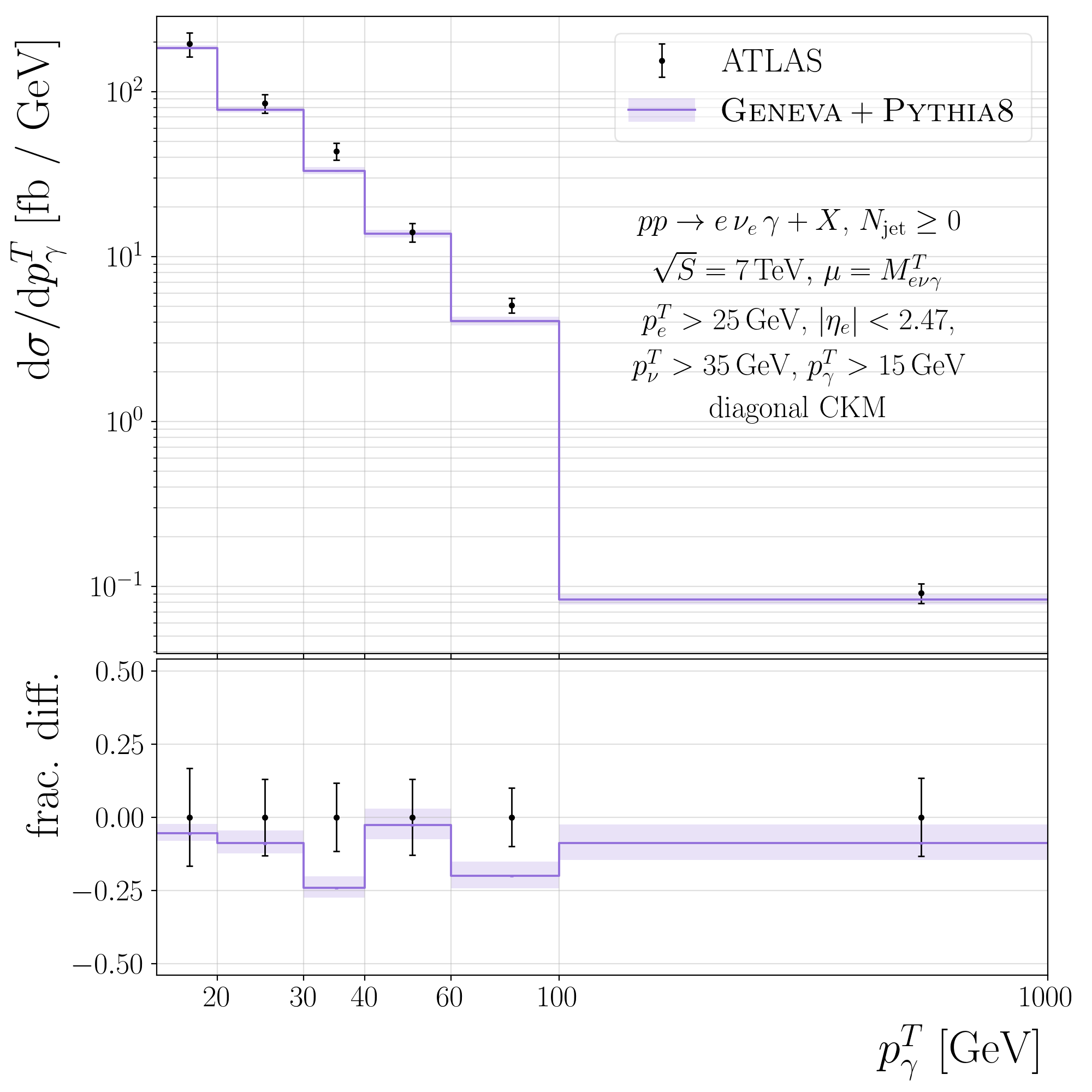}%
  \end{subfigure}
  \hspace*{\hspacebetweentwoplots}
  \begin{subfigure}[b]{\rescaletwoplots}
    \includegraphics[width=\textwidth]{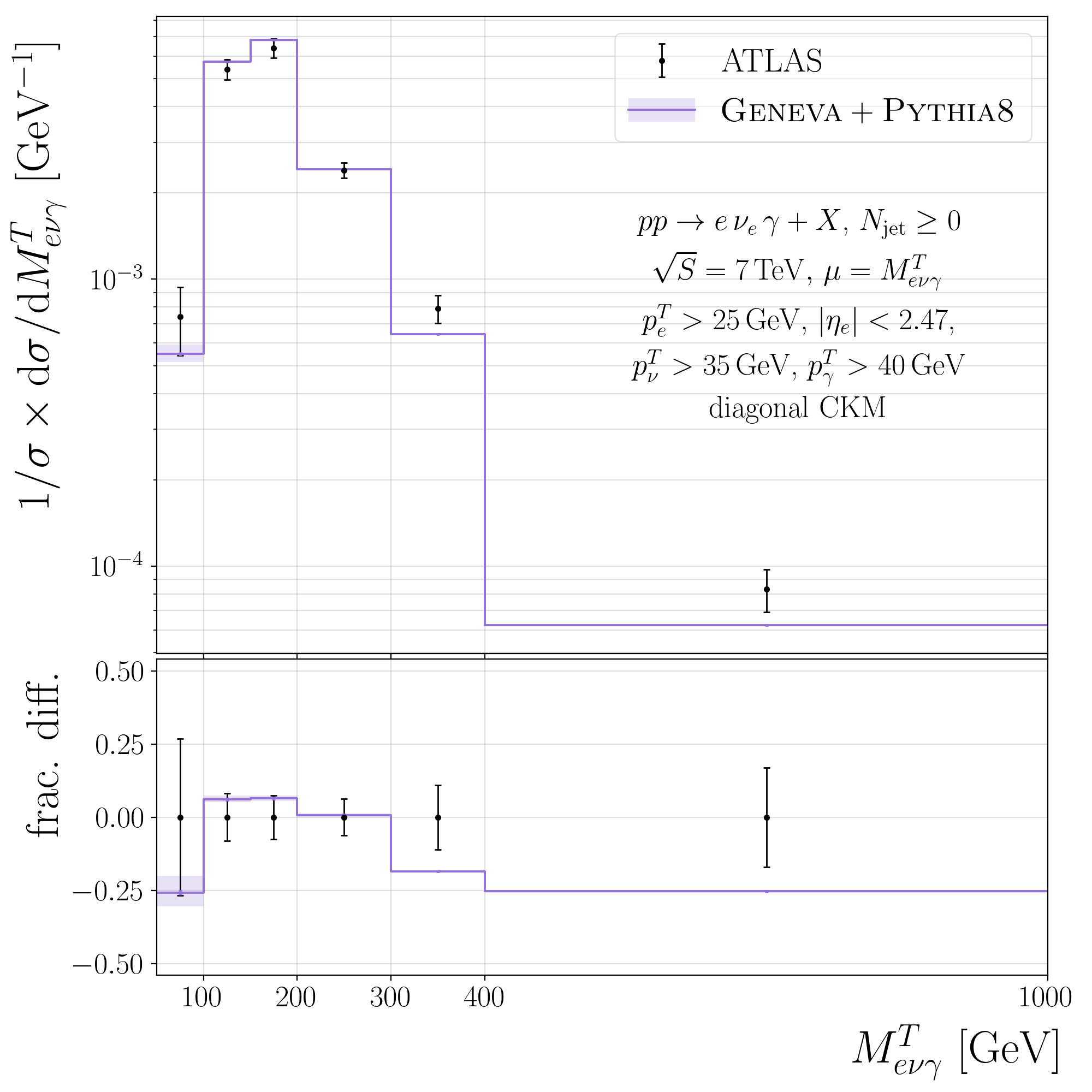}%
  \end{subfigure}
  \vspace{\spaceabovefigurecaption}
  \caption{Comparison between \geneva predictions and ATLAS measurements at 7 TeV (4.6 fb$^{-1}$). Selection cuts, bin widths and observable definitions are defined in the original ATLAS~\cite{Aad:2013izg} publication.}
  \label{fig:gvavsATLAS}
\end{figure*}

Finally, in \fig{gvavsATLAS} we compare the \geneva predictions to data collected by the ATLAS experiment at the LHC in $\sqrt{S}=7$ TeV collisions, with a total integrated luminosity of 4.6 fb$^{-1}$~\cite{Aad:2013izg}. We generate events with a set of isolation cuts which are looser than those used in the experimental analysis, and then pass the resulting events through \rivet~\cite{Bierlich:2019rhm} -- this amounts to using a hybrid-type photon isolation. The sum over events with both signs of the intermediate $W$ boson is considered. We show the distribution of the transverse momentum of the photon $p^T_\gamma$, and the normalised distribution of the transverse mass of the colour singlet system  $M^T_{e\nu\gamma}$, the latter with a harder cut on the photon of $p^T_\gamma>40$ GeV. The agreement with data is reasonably good and of a similar quality to the NNLO results provided by \matrix in Ref.~\cite{Grazzini:2015nwa}. In particular, for the case of the $M^T_{e\nu\gamma}$ distribution we notice that the predictions tend to undershoot the data at high values. We anticipate that the inclusion of electroweak corrections would improve the agreement in this region, where the contribution due to photon-initiated processes is known to be large (though partially mitigated by correspondingly large and negative corrections to the $q\bar{q}$ channel)~\cite{Denner:2014bna}.

\section{Conclusions}
\label{sec:conc}

In this work we have presented the first event generator at NNLO accuracy for the diboson production process $pp\to \ell\nu_{\ell}\gamma$ matched to a parton shower. This has been implemented in the \geneva framework, which matches a resummed calculation at NNLL$'$ in the resolution variable $\Tau_0$ to fixed order predictions at NNLO and allows the resulting events to be passed to \pythiaEight. We validated the NNLO accuracy of our predictions for inclusive observables by comparing with the fixed order code \matrix, finding that the kinematic effects of missing nonsingular power corrections on differential distributions are mild at worst. We then examined the effect of the shower on the $\Tau_0$ distribution, finding that the NNLL$'$ accuracy was numerically well-preserved in the peak region, and observed that the effect of hadronisation is limited to a shift at low values where nonperturbative effects are particularly relevant.

Lastly, we compared to data collected at the 7 TeV LHC by the ATLAS experiment in 2011. We found good agreement, consistent with previous studies at fixed order in QCD, with small deviations seen in regions where EW effects are likely to be important.

It would be interesting to consider the inclusion of these electroweak corrections to this process, especially given their numerical relevance at the 13 TeV LHC~\cite{Campbell:2021mlr}. A study of the differences with respect to $Z\gamma$ production and the effect of the inclusion of a non-diagonal CKM matrix may also prove interesting -- we leave these issues to a future work.

\section*{Acknowledgements}
\label{sec:Acknowledgements}
We thank our collaborators Simone Alioli, Alessandro Broggio, Alessandro Gavardi, Stefan Kallweit, and Davide Napoletano for useful discussions. We thank Stefan Kallweit for help with \matrix and for providing the necessary \munich libraries. The work of MAL and RN is supported by the ERC Starting Grant REINVENT-714788. TC is supported by the Science and Technology Facilities Council (STFC) via grant awards ST/P000274/1 and ST/T000856/1. We acknowledge the CINECA award under the ISCRA initiative for the availability of the high performance computing resources needed for this work.

\appendix

\section{Nonsingular power corrections}
\label{app:power_corrections}

\begin{figure}[t!]
    \includegraphics[width=0.45\textwidth]{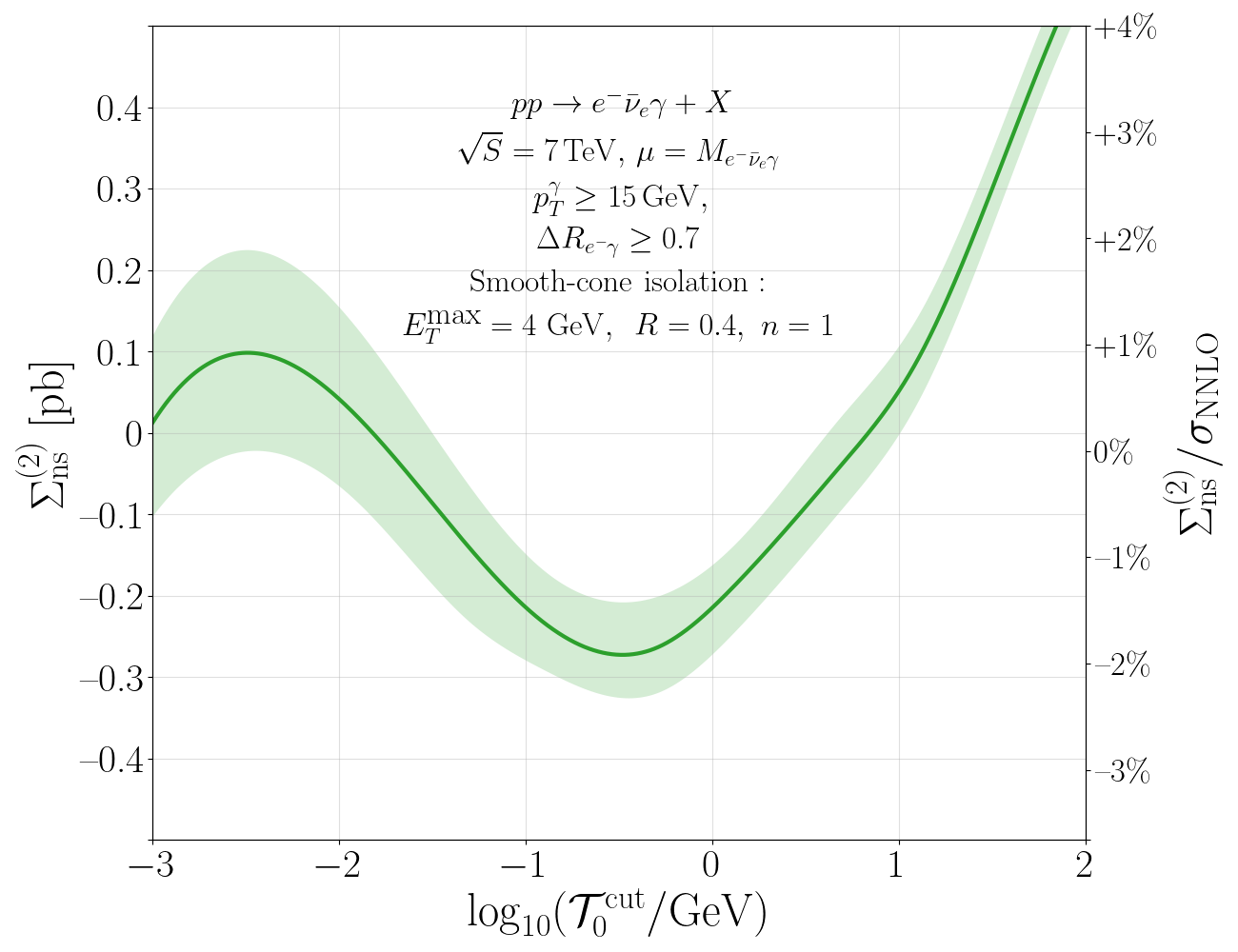}%
  \vspace{\spaceabovefigurecaption}
  \caption{Size of the neglected power corrections
    $\Sigma^{(2)}_{\mathrm{ns}}$ as a function of the resolution
    parameter $\Tau_0^\cut$. The target result $\sigma_{\rm NNLO}$ is
    the fixed-order cross section as computed by \matrix. The error band
    shows the size of the numerical fluctuations. On the right axis we also show the relative size
of the corrections as a fraction of the \matrix NNLO cross section.
    \label{fig:power_corrections}
  }
\end{figure}

The contribution to the $0$-jet bin below the resolution cut-off $\Tau_0^\cut$ (\eq{0full}) would require
a local NNLO subtraction in order to be properly implemented. Instead, we substitute the expression with
\begin{align} \label{eq:0masterfultilde}
\frac{\dsigtildeMCz}{\df\Phi_0}(\Tau_0^\cut)
&= \frac{\df\sigma^{\rm NNLL'}}{\df\Phi_0}(\Tau_0^\cut)
- \biggl[\frac{\df\sigma^{\rm NNLL'}}{\df\Phi_0}(\Tau_0^\cut) \biggr]_{\rm NLO_0}
\nn \\ & \quad
+ (B_0 + V_0)(\Phi_0)\, \thetaPSiso(\Phi_0) 
\nn \\ & \quad
+ \int \! \frac{\df \Phi_1}{\df \Phi_0}\, B_1(\Phi_1)\, \thetaPSiso (\Phi_1)\nn \\
&\qquad\times\,\thetaProj(\widetilde{\Phi}_0)\, \theta\big(\Tau_0(\Phi_1) < \Tau_0^\cut\big)\,
\,.
\end{align}
and so neglect the contribution
\begin{align} \label{eq:nonsingcum}
\frac{\df\Sigma^{(2)}_{\mathrm{ns}}}{\df\Phi_0}(\Tau_0^\cut)
&= \frac{\dsigMC_0}{\df\Phi_0}(\Tau_0^\cut) - \frac{\dsigtildeMCz}{\df\Phi_0}(\Tau_0^\cut)
\, .
\end{align}
These neglected terms are power corrections of $\ord{\alpha_s^2}$. They therefore contribute at the same
order as the power corrections which arise from evaluating higher multiplicity configurations on projected
phase space points, which is an issue common to any method of event generation.
The integral of all these terms is shown in \fig{power_corrections} as a
function of $\Tau_0^\cut$.
We observe that the behaviour is indeed nonsingular so that the size of the corrections is reduced at smaller values of
$\Tau_0^\cut$.
For the purposes of this Letter we set
\begin{equation}
  \Tau_0^\cut = 0.5 \GeV\,.
\end{equation}
This choice is the smallest value of $\Tau_0^\cut$ for which we could maintain a positive $\sigma_0^\mathrm{MC}$.
Despite the larger size of power corrections at this value, we observe that the effect on the total cross section is still relatively small, at $< 2\%$.
We are able to recover the effect of these missing contributions on the total cross section and its variations by reweighting the events below
$\Tau_0^\cut$. As with any event generator, however, we miss the $\ord{\alpha_s^2}$ nonsingular kinematic dependence of the power corrections
for this class of events. The comparisons in \fig{gvavsmatrix} attest to the fact that this omission does not significantly affect our predictions
for differential distributions.

\bibliographystyle{elsarticle-num}
\bibliography{geneva}

\begin{thebibliography}{10}
\bibitem{munich}
{\sc Munich} is the abbreviation of ``MUlti-chaNnel Integrator at Swiss (CH)
  precision''---an automated parton level NLO generator by S.~Kallweit. In
  preparation.
\end{thebibliography}

\end{document}